\documentclass{article}
\usepackage{spconf,amsmath,graphicx,url,subfig, cite,booktabs}
\usepackage{amsfonts,amssymb}
\usepackage{xcolor}
\title{Expressive-VC: Highly expressive Voice Conversion with Attention Fusion of Bottleneck and Perturbation Features}

\name{
\begin{tabular}{c}
\it Ziqian Ning$^1$, Qicong Xie$^1$, Pengcheng Zhu$^2$, Zhichao Wang$^1$, Liumeng Xue$^1$, Jixun Yao$^1$\thanks{* Corresponding author.}, \\
\it Lei Xie$^{1*}$, Mengxiao Bi$^2$
\end{tabular}
}
\address{
  $^1$Audio, Speech and Language Processing Group (ASLP@NPU), School of Computer Science, \\ Northwestern Polytechnical University, Xi'an, China\\
  $^2$Fuxi AI Lab, NetEase Inc., Hangzhou, China
  }

\begin{document}
\ninept
\maketitle
\begin{abstract}
\vspace{-6pt}
Voice conversion for highly expressive speech is challenging. Current approaches struggle with the balancing between speaker similarity, intelligibility and expressiveness.
To address this problem, we propose \textit{Expressive-VC}, a novel end-to-end voice conversion framework that leverages advantages from both neural bottleneck feature (BNF) approach and information perturbation approach.
 Specifically, we use a BNF encoder and a Perturbed-Wav encoder to form a content extractor to learn \textit{linguistic} and \textit{para-linguistic} features respectively, where BNFs come from a robust pre-trained ASR model and the perturbed wave becomes speaker-irrelevant after signal perturbation. We further fuse the linguistic and para-linguistic features through an attention mechanism, where speaker-dependent prosody features are adopted as the attention query, which result from a prosody encoder with target speaker embedding and normalized pitch and energy of source speech as input. Finally the decoder consumes the integrated features and the speaker-dependent prosody feature to generate the converted speech. Experiments demonstrate that Expressive-VC is superior to several state-of-the-art systems, achieving both high expressiveness captured from the source speech and high speaker similarity with the target speaker; meanwhile intelligibility is well maintained.

% Expressive-VC. based on feature fusion is proposed for highly expressive VC.

\end{abstract}
% Specifically, with the robust disentanglement performance of linguistic information extracted from automatic speech recognition (ASR) and the high reconstruction quality of perturbation-based methods, an attention-based fusion module is proposed to combine the advantages of these two methods. With the guidance of prosodic information, the fusion content generated by the fusion module obtains linguistic information from bottleneck feature of ASR and fetches additional para-linguistic and non-verbal information from perturbed waveform. Extensive experiments show that the proposed method can achieve superior performance for expressive VC on non-expressive, expressive, and even non-verbal scenes.
\vspace{-4pt}
%
\iffalse
\begin{figure*}[htbp]
    \centering
    % \includegraphics[scale=0.8]{figs/overall.svg}
    \includegraphics[scale=0.35]{figs/overall.pdf}\vspace{-0.3cm}
    % \includegraphics[scale=0.67]{IEEEtran/fig/overall-s.pdf}\vspace{-0.1cm}
    \caption{The overall architecture for Expressive-VC}\vspace{-0.3cm}
    \label{fig:model}
\end{figure*}
\fi

\begin{figure*}[!htbp]
    % \label{fig:pitch}
  \begin{minipage}[t]{0.33\linewidth}
    \centering    
% \subfloat[content encoder]{\includegraphics[trim=20 0 20 0,scale=0.7]{IEEEtran/fig/content.pdf}}
\subfloat[Expressive-VC]{\includegraphics[clip,scale=0.38]{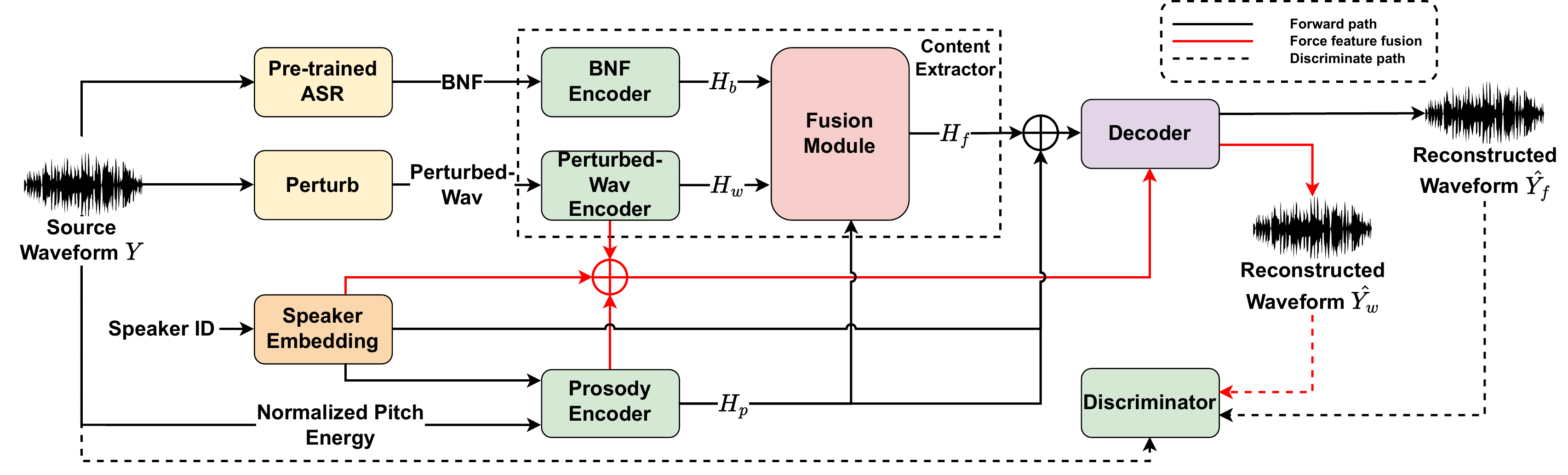}}
    % \label{fig:side:a}
  \end{minipage}%
  \begin{minipage}[t]{1.1\linewidth}
    \centering
\subfloat[Fusion Module]{\includegraphics[clip,scale=0.45]{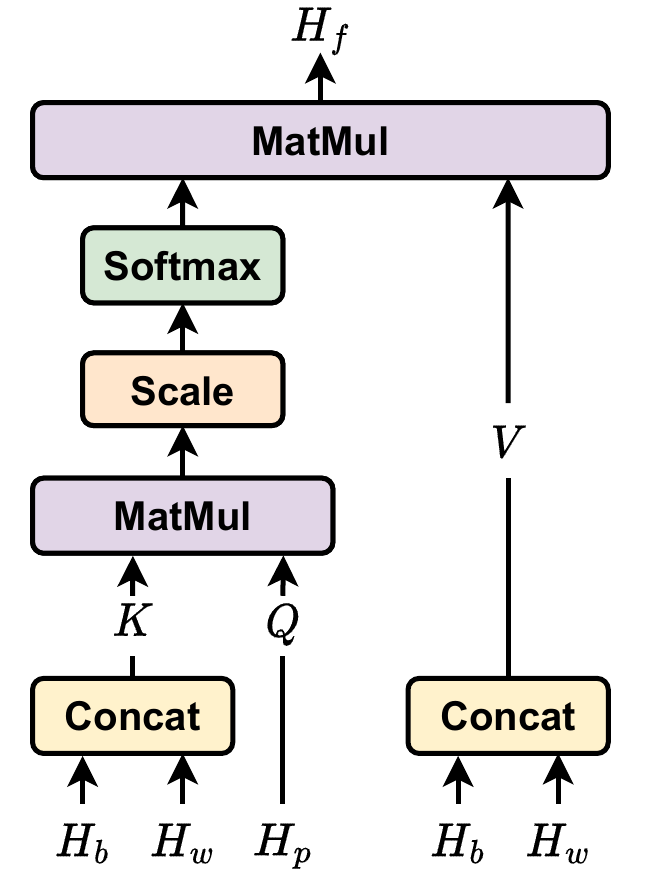}}
    % \caption{cap b}
    % \label{fig:side:b}
  \end{minipage}
  \vspace{-0.26cm} 
     \caption{The architecture of (a) Expressive-VC and (b) fusion module.}
     \vspace{-8pt}
    \label{fig:model}\vspace{-0.45cm} 
\end{figure*}

\vspace{-14pt}
\section{Introduction}
\label{sec:intro}
\vspace{-8pt}

The information conveyed in human speech can be roughly categorized into linguistic, para-linguistic and non-linguistic aspects, representing language, rhythmic-emotional and speaker identity respectively~\cite{yamashita2013review}.
Moreover, non-verbal sounds, such as breathing, laughing and crying, are also essential in speech communication.
Voice conversion (VC) is a technique that alters speaker-related information in a given speech to make it sound like another speaker while preserving the rest aspects of speech~\cite{DBLP:journals/taslp/SismanYKL21}, ideally including linguistic, para-linguistic and even non-verbal aspects. It has been widely used in scenarios such as personalized speech synthesis~\cite{DBLP:conf/icml/QianZCYH19} and privacy protection~\cite{DBLP:journals/csl/TomashenkoWVPSN22}. With the advances of deep learning, voice conversion has drawn much attention in more challenging scenarios such as movie dubbing with highly expressive speech contents.

In voice conversion, the essential task is to decouple the speaker-relevant and -irrelevant information from the source audio and transfer the aforementioned speaker-irrelevant information to the target speaker. This is not a trivial task as those components are highly entangled in speech. 

%Generally speaking, this challenging disentanglement problem can be addressed in two different ways~\cite{DBLP:conf/icassp/YeZSHRLL22}.

%there are two directions to approach this challenging problem --  explicit disentanglement and implicit disentanglement~\cite{DBLP:conf/icassp/YeZSHRLL22}.

Some disentanglement approaches rely on the fine-grained design of the voice conversion model itself, including adopting specific modules, losses, or learning algorithms in the voice conversion model to constrain the learned feature to represent either linguistic content or speaker identity. For example, vector quantization (VQ)~\cite{DBLP:journals/tetci/HuangLHLPTW20,DBLP:conf/interspeech/WangDYCLM21,DBLP:conf/interspeech/DingG19}, adaptive instance normalization~\cite{DBLP:conf/icassp/ChenWWL21} and gradient reversal layer (GRL)~\cite{DBLP:conf/icassp/LiTYWXSM21} are adopted to obtain relatively pure linguistic-related feature and remove speaker-related information in the learned feature~\cite{DBLP:journals/tetci/HuangLHLPTW20,DBLP:conf/interspeech/WangDYCLM21,DBLP:conf/interspeech/DingG19,DBLP:conf/icassp/ChenWWL21,DBLP:conf/icassp/LiTYWXSM21}. Furthermore, mutual information (MI) loss can be adopted to minimize the correlation between the speaker and linguistic information in the learned feature~\cite{DBLP:conf/interspeech/WangDYCLM21,DBLP:conf/interspeech/YangTZ0SWCTZWM22}. 
%Explicit disentanglement
These approaches often face a dilemma in real applications -- to maintain reasonable target speaker similarity, there has to be an empirical compensation on the transferred expressiveness; otherwise, the source speaker's timbre may leak to the target speaker, making converted speech sound somehow
like uttered by the source speaker or a mixture of both.

Some prior disentanglement actions can be conducted outside or before the voice conversion model. Taking the decoupled features as input, the VC model can make further disentanglement more easily. In this direction, Phonetic PosteriorGrams (PPGs) and neural network bottleneck features (BNFs) are usually adopted to help the disentanglement. Specifically, BNFs are a set of activation of nodes over time from a neural network bottleneck layer,
while PPGs are obtained by stacking a sequence of phonetic posterior probabilistic vectors from the neural network. Both BNFs and PPGs, usually obtained from a well-trained acoustic model in an automatic speech recognition (ASR) system, are proven to be linguistic-rich, speaker-independent and noise-robust. Thus using the PPGs/BNFs as intermediate representation, the voice conversion process is factorized as a speech-to-BNFs/PPGs module and a BNFs/PPGs-to-speech module, or the 
so-called recognition-synthesis framework~\cite{DBLP:conf/icmcs/SunLWKM16,DBLP:conf/icassp/LiTYWXSM21,DBLP:conf/icassp/ZhaoLSWKTM22,DBLP:conf/interspeech/TianC019}. In this way, the linguistic information embedded in the source speech can be transferred stably to the target speech. As containing mostly the linguistic information in BNFs/PPGs, the converted speech apparently loses the expressiveness of the source speech to a large extent. The use of extra prosody features, such as pitch, can be a remedy to this problem~\cite{DBLP:conf/interspeech/LiuCKHL00M20,DBLP:conf/interspeech/WangZYLDXGCL21,DBLP:conf/iscslp/LianZWLT21}.

%the specific information-related feature is obtained at the stage of data pre-processing rather than conversion model training. For example, bottleneck features (BNF), which is extracted from a pre-trained automatic speech recognition (ASR) model on a large amount of data, can be used as a stable linguistic content feature, achieving more stable pronunciation in converted speech~\cite{DBLP:conf/iscslp/WangGWYGCLXL21, DBLP:conf/iccv/HuangB17}. The BNF extracted from speech is highly related to linguistic content for that the training target of the ASR models is text transcript, so information beyond the linguistic content in speech is discarded, leading to the lack of expressiveness in the converted speech.

Recently, information perturbation has been introduced to remove speaker timbre in prior for voice conversion ~\cite{DBLP:conf/nips/ChoiLKLHL21}. 
% The basic idea of information perturbation is to perturb all the unwanted information in the speech by signal processing in advance so that training the neural network model to not extract the undesired attributes. 
The basic idea of information perturbation is to process all the unwanted information in the speech by signal processing beforehand which can make the neural network learn the essential information effectively. 
Specifically, information perturbation is adopted to remove speaker-related information in the source speech and thus the linguistic information is subsequently modeled by a content encoder~\cite{DBLP:journals/corr/abs-2206-07569}. Beyond the linguistic information, para-linguistic, e.g., emotional information, can also be preserved after speaker perturbation in the speech~\cite{DBLP:journals/spl/LeiYZXS22}.  In this way, the voice model no longer suffers from the aforementioned trade-off between speaker similarity and expressiveness. Thus it is promising for information perturbation based method to transfer all the expressions in the source speech to the target speaker while maintaining good similarity with the target speaker. However, as the perturbation parameters are empirically selected, this kind of method may lack robustness in intelligibility and quality in the converted speech.

To achieve voice conversion for highly expressive source speech, in this paper, we propose \textit{Expressive-VC}, a novel end-to-end voice conversion framework that leverages advantages from both neural bottleneck feature approach and information perturbation approach.
 Specifically, we use a BNF encoder and a Perturbed-Wav encoder to form a content extractor to learn \textit{linguistic} and \textit{para-linguistic} features respectively, where BNFs come from a robust pre-trained ASR model and the perturbed wave is considered to be speaker-irrelevant and after signal perturbation. We further fuse the linguistic and para-linguistic features through a scaled dot-product attention mechanism, where speaker-dependent prosody features are adopted as the attention query, which result from a prosody encoder with target speaker embedding and normalized pitch and energy of source speech as input. Finally the decoder consumes the fused feature and speaker-dependent prosody feature to generate the converted speech. Extensive experiments demonstrate that Expressive-VC is superior to several competitive systems, achieving both high expressiveness captured from the source speech and high speaker similarity with the target speaker; meanwhile intelligibility is well maintained.
\vspace{-12pt}
\section{Proposed Approach}
\label{sec:format}
\vspace{-8pt}

% As shown in Figure~\ref{fig:model}, the proposed model architecture consists of five components: 
% (i) A content encoder to extract semantics and emotions, expressiveness, and para-linguistic information from the waveform after perturb
% (ii) A BNFs encoder to further extract content information from BNFs
% (iii) A pitch encoder to generate speaker-related pitch from normalized pitch
% (iv) A fusion module to fuse BNFs and LFs
% (v) A waveform generator for high-quality waveform generation

% Here using one or two sentences introducing the general idea...
To perform expressive voice conversion, we design our Expressive-VC system by fusing robust linguistic information embedded in the bottleneck feature (BNF) and rich para-linguistic information contained in the speaker-attribute-perturbed wave (Perturbed-wav). As shown in Fig.~\ref{fig:model}(a), Expressive-VC is based on an encoder-decoder architecture, mainly consisting of four components -- content extractor, prosody encoder, decoder and discriminator. 
% robust disentanglement ability of BNF-based methods and high reconstruction quality of perturbation-based methods. The proposed approach and training strategy will be detailedly introduced in this section.

\vspace{-12pt}
\subsection{Content Extractor}
\vspace{-6pt}
The content extractor is composed of a BNF encoder, a Perturbed-Wav encoder and a fusion module. The two encoders are designed to learn linguistic and para-linguistic features from source speech, respectively. Subsequently the fusion module fuses the two types of features for better expressivity and reasonable intelligibility in the converted speech.
% high-quality and high-expressiveness voice conversion.
%into comprehensive linguistic information 
%The BNF encoder takes BNF as input and output linguistic information $H_{b}$ in the source speech, and P-wav encoder takes P-wav as input to generate para-linguistic information  $H_{w}$ of the source speech. The fusion module is adopted to effectively combine $H_{b}$ and $H_{w}$ and generate the comprehensive linguistic information $H_{f}$. 

% The prosody encoder provides speaker-related prosody representation $H_{p}$. The decoder reconstructs waveform from content representation, speaker embedding, and prosody representation. During training process, discriminators including multi-period discriminator (MPD), multi-scale discriminator (MSD), and multi-resolution spectrogram discriminator provide constraints from temporal and frequency aspects.

% With the guidance of prosody information, the fusion module in content extractor can generate comprehensive content information from the outputs of BNF encoder and P-wav encoder. 

% Specifically, content encoder and prosody encoder extract linguistic content and para-linguistic information from BNF and perturbed waveform, respectively. By fusion these two kinds of representations, the fusion module can obtain comprehensive content information. The speaker encoder learns to extract speaker-dependent prosody representations. Taking content, prosody, and speaker timbre as input, the decoder reconstructs the waveform. The proposed attention-based fusion method and training strategy will be detailedly introduced in this section.

\textbf{BNF \& Perturbed-Wav Encoders}
The BNF encoder takes BNFs as input and output linguistic embedding $H_{b}$ for the source speech, and the Perturbed-Wav encoder takes the perturbed wave as input to generate para-linguistic-related embedding $H_{w}$ of the source speech. $H_{b}$ and $H_{w}$ $ \in \mathbb{R}^{T \times F}$, where $T$ represents the sequence length and $F$ is the dimension of the embeddings. The BNFs are extracted from the source waveform $Y$ by a pre-trained ASR model. The perturbed wave is the waveform perturbed by three signal processing functions: pitch randomization (pr), formant shifting (fs), and random frequency shaping using a parametric equalizer (peq). The pitch randomization function shifts the pitch and scales its range, and the formant shifting function also shifts the formants, which encourages to change the speaker timbre in the source waveform. By modifying the energy of different frequency bands, the parametric equalizer function further removes the speaker-relevant information. In summary, the speaker perturbation process on the source waveform $Y$ can be simply described as
\vspace{-6pt}
\begin{equation}
% Y^{\prime} = pr(fs(peq(Y)))
Perturbed\mbox{-}wav = pr(fs(peq(Y))),
\vspace{-6pt}
\end{equation}
where the perturbed wave is regarded as speaker irrelevant while the general linguistic and para-linguistic pattern are maintained. 
\begin{table*}[!htbp]
\caption{Comparison between Expressive-VC (with ablation), BNF-VC, Perturb-VC, and AGAIN-VC, in terms of speaker similarity MOS (SMOS) and naturalness MOS (NMOS) with confidence intervals of $95\%$ under 3 voice conversion scenarios. Character Error Rate (CER) is also calculated for intelligibility measure. CER for source speech is 6.1\% (non-expressive) and 10.3\% (expressive).}
 \label{tab:mos}
 \vspace{-0.2cm}
\setlength{\tabcolsep}{3mm}
 \centering
 \resizebox{\linewidth}{!}{
 %\footnotesize
% \usepackage{booktabs}
\begin{tabular}{@{}l|ccr|ccr|cc|cc@{}}
\toprule
                                                                    & \multicolumn{3}{c|}{Non-expressive}                                                        & \multicolumn{3}{c|}{Expressive}                                                            & \multicolumn{2}{c|}{Non-verbal}                                                     & \multicolumn{2}{c}{Overall}                                                         \\ \midrule
                                                                    & NMOS $\uparrow$                                    & SMOS $\uparrow$                                    & CER $\downarrow$ & NMOS $\uparrow$                                    & SMOS $\uparrow$                                    & CER $\downarrow$ & NMOS  $\uparrow$                                   & SMOS $\uparrow$                                    & NMOS $\uparrow$                                    & SMOS $\uparrow$                                    \\ \midrule
BNF-VC                                                              & 3.97$\pm$0.034                           & \textbf{3.91$\pm$0.045} & \textbf{7.3}  & 3.81$\pm$0.043                           & 3.59$\pm$0.057                           & \textbf{11.5} & 3.75$\pm$0.022                           & 3.59$\pm$0.050                           & 3.84$\pm$0.042                           & 3.70$\pm$0.032                           \\
Perturb-VC                                                          & 3.67$\pm$0.040                           & 3.32$\pm$0.037                           & 9.7  & 3.55$\pm$0.048                           & 3.66$\pm$0.049                           & 16.9 & 3.76$\pm$0.050                           & 3.22$\pm$0.065                           & 3.66$\pm$0.050                           & 3.41$\pm$0.030                           \\
AGAIN-VC                                                            & 2.81$\pm$0.037                           & 3.16$\pm$0.048                           & 14.1 & 2.80$\pm$0.034                           & 2.96$\pm$0.028                           & 22.2 & 2.66$\pm$0.041                           & 2.77$\pm$0.029                           & 2.76$\pm$0.036                           & 2.96$\pm$0.046                           \\ \midrule
Expressive-VC                                                       & \textbf{4.00$\pm$0.049} & 3.81$\pm$0.035                           & 8.7  & \textbf{4.05$\pm$0.042} & \textbf{3.78$\pm$0.032} & 11.8 & \textbf{4.06$\pm$0.041} & \textbf{3.83$\pm$0.035} & \textbf{4.04$\pm$0.034} & \textbf{3.81$\pm$0.040} \\
\hspace{1em}-Fusion Module                         & 3.57$\pm$0.050                           & 3.62$\pm$0.034                           & 9.2  & 3.68$\pm$0.048                           & 3.59$\pm$0.028                           & 14.9  & 3.90$\pm$0.032                           & 3.61$\pm$0.028                           & 3.71$\pm$0.042                           & 3.61$\pm$0.043                           \\
\hspace{1em}-$\mathcal{L}_{G_c},\mathcal{L}_{D_c}$ & 3.88$\pm$0.033                           & 3.72$\pm$0.037                           & 8.8  & 3.79$\pm$0.034                           & 3.75$\pm$0.040                           & 13.5  & 3.37$\pm$0.058                           & 3.41$\pm$0.024                           & 3.68$\pm$0.032                           & 3.63$\pm$0.049                           \\
\hspace{1em}-Speed Aug                             & 3.71$\pm$0.042                           & 3.71$\pm$0.034                           & 10.4  & 3.96$\pm$0.048                           & 3.75$\pm$0.029                           & 18.1  & 3.46$\pm$0.034                           & 3.27$\pm$0.026                           & 3.71$\pm$0.034                           & 3.58$\pm$0.034                           \\ \bottomrule
\end{tabular}
 }
\vspace{-18pt}
\end{table*}
% \subsection{Attention-based Feature Fusion} \label{attention-section}
\textbf{Feature Fusion Module} 
% We assume that BNFs contain robust speaker-independent content information, while LFs have a more general pronunciation representation which is able to retain more expressive and para-linguistic information. An intuitive idea to fuse these two features is using concatenation or addition, but content information is also present in LFs, which means that the fused features will have a large amount of redundancy, increasing the difficulty of modeling in the decoder module. The ideal fusion should be to obtain the main content information from BNFs and the complementary expressive and paralinguistic information from LFs. 
% By using prosody features, including energy and f0, we can make a judgment about the characteristics of the current speech, whether it is spoken in a normal tone, a highly expressive way of speaking, or para-linguistic information that is difficult to represent the content. In turn, the proportion of the two features in the fused features can be dynamically adjusted to fully express the information the speech contains while being able to minimize the information redundancy. 
Obtaining robust and rich content representation including both linguistic and para-linguistic information from source speech is essential in VC tasks. As discussed earlier, BNFs are considered to be linguistic-rich but lose most of the expressivity in the speech. By contrast, the embedding extracted from the speaker-perturbed wave may contain rich expressive aspects of speech. An intuitive idea to combine both features by simple addition or concatenation. However, we believe that the fusion should be done dynamically because the contributions from linguistic and para-linguistic aspects vary through time. For example, non-verbal sounds such as breathing may have low contribution from the BNF embedding but high contribution from the Perturbed-wav embedding. Note that those non-verbal sounds are not explicitly modeled in an acoustic model of ASR.

%BNF extracted from the ASR model is widely used to achieve this goal. But with distorted para-linguistic information, BNF usually leads to low speaking style similarity between the source and converted speeches, especially in highly expressive application scenarios. To address this problem, an intuitive idea is to combine the linguistic information of BNF and para-linguistic information extracted from perturbed waveform by addition or concatenation~\cite{DBLP:conf/interspeech/LiuCKHL00M20,DBLP:conf/iscslp/LianZWLT21,DBLP:conf/interspeech/WangZYLDXGCL21}. However, simple addition or concatenation cannot accurately correct the distorted para-linguistic information, resulting in potential unstable issue.
% which potentially causes unstable performance. 

% Alternatively, the ideal fusion should be to mainly obtain linguistic-content information from BNF and utilize the para-linguistic and even non-verbal information extracted by the P-wav encoder when distortion occurs in BNF.

To realize dynamic fusion, we propose an attention-based fusion module to effectively combine the linguistic feature $H_{b}$ and para-linguistic feature $H_{w}$. As shown in Fig.~\ref{fig:model} (b), the concatenation result of $H_{b}$ and $H_{w}$ is used as key $K \in \mathbb{R}^{T \times 2 \times F}$  and value $V \in \mathbb{R}^{T \times 2 \times F}$. The output of prosody encoder (described in Section 2.2) $H_{p}$ is used as query $ Q \in \mathbb{R}^{T \times F \times 1}$ to integrate the linguistic and para-linguistic information in $H_{b}$ and $H_{w}$. In other words, we use the general prosody pattern of the source speech (source speaker timbre removed) to weight the fusion of the two branches. Following the attention mechanism in~\cite{DBLP:conf/nips/VaswaniSPUJGKP17}, we use the scaled dot-product operation as the similarity measure. The whole process of the proposed fusion module can be described as:
% To guide the process of attention, the output of prosody encoder $H_{p} \in \mathbb{R}^{T \times F}$, generated from energy and pitch, is used to indicate the integrity of content information in $H_{b}$ and $H_{w}$. Specifically, in our attention-based method, the concatenation of $H_{b}$ and $H_{w}$ is used as key $K$ and value $V$ and $H_{p}$ is used as query $Q$. Following the method in scaled dot-product attention~\cite{DBLP:conf/nips/VaswaniSPUJGKP17}, scaled dot-product is used as a similarity measure method. The whole process of the proposed fusion module can be described as:
\vspace{-8pt}
% \begin{equation}
% \begin{split}
% % \centering
% K=V=concat(unsqueeze(H_{b}),unsqueeze(H_{w}))
% \vspace{-4pt}
% \end{split}
% \end{equation}
% \begin{equation}
% \centering
% Q=unsqueeze(H_{p})
% \vspace{-4pt}
% \end{equation}
% \begin{equation}
% attention(Q, K) = softmax(\frac{QK}{\sqrt{F}}),
% \vspace{-4pt}
% \end{equation}
% \begin{equation}
% \begin{split}
% H_{f}=suqeeze(attention(Q, K) V)
% \end{split}
% \end{equation}
\begin{equation}
\begin{split}
% \centering
K=V=concat(H_{b},H_{w})
\end{split}
\vspace{-8pt}
\end{equation}
%\vspace{-8pt}
\begin{equation}
\centering
%\vspace{-12pt}
Q=H_{p}
%\vspace{-8pt}
\end{equation}
%\vspace{-8pt}
\begin{equation}
attention(Q, K) = softmax(\frac{QK}{\sqrt{F}}),
%\vspace{-4pt}
\end{equation}
\begin{equation}
\begin{split}
H_{f}=attention(Q, K) V
\end{split}
\end{equation}
where  $H_{f} \in \mathbb{R}^{T \times F}$ is the output of the fusion module.

\vspace{-8pt}
\subsection{Prosody Encoder}
\vspace{-4pt}
To better preserve the prosody in the source speech and obtain high speaker similarity with the target speech, inspired by~\cite{DBLP:journals/corr/abs-2206-07569}, a prosody encoder is used to learn the speaker-related prosody representation. First, pitch ($f_0$) and energy ($e$) are extracted from the source speech $Y$ and then z-score normalization is performed on pitch to remove the source speaker's timbre, yielding a speaker-independent prosody. Then the conditional layer normalization (CLN)~\cite{DBLP:conf/iclr/Chen0LLQZL21} is adopted to generate the target speaker-related prosody feature $H_p$ by using the target speaker embedding as the conditional information:
\vspace{-6pt}
\begin{equation}
\begin{split}
    H_p &= concat(\gamma \frac{f_0-\mu(f_0)}{\sigma(f_0)} + \beta, e)
\end{split}
\vspace{-6pt}
\end{equation}
where $\gamma$ and $\beta$ are scale and bias vectors about speaker embedding, while $\mu(f_0)$ and $\sigma(f_0)$ stand for utterance level mean and variance of $f_0$ respectively. $H_p$ has two usages -- one is used as the attention query in the aforementioned fusion module while another is fed to the decoder together with the content extractor output. 

% \(\mu\) and \(\sigma\) are the mean and variance of pitch \(p\), and \(e\) denotes energy.

\iffalse
\begin{equation}
\begin{split}
    p^\prime &= \frac{p-\mu}{\sigma} \\
    H_p &= concat(q^\prime, e)
\end{split}
\end{equation}
\fi

\iffalse
\subsection{Prosody Encoder}
To better preserve the prosodic information in the source speech and obtain high speaker similarity with the target speech, inspired by~\cite{DBLP:journals/corr/abs-2206-07569}, a prosody encoder is used to learn the speaker-related prosody representation. First, we normalize the fundamental frequency (normalized pitch) and short-term average amplitude (energy) extracted from the source speech to obtain a speaker-independent prosody. Then the conditional layer normalization (CLN)~\cite{DBLP:conf/iclr/Chen0LLQZL21} is adopted to generate speaker-related pr
\fi
% the speaker embedding is used as the conditional information to output a scale and bias
%The prosody encoder uses speaker embedding and prosodic features, 
% including normalized fundamental frequency (normalized pitch) and short-term average amplitude (energy), to obtain speaker-related prosody representation.

\vspace{-8pt}
\subsection{Decoder and Discriminator}
\vspace{-4pt}
With the input of source speech and target speaker identity, our proposed model directly reconstructs waveform without an explicit vocoder. Our decoder follows HiFi-GAN~\cite{DBLP:conf/nips/KongKB20} using multiple discriminators for adversarial training, including multi-period discriminator (MPD), multi-scale discriminator (MSD), and multi-resolution spectrogram discriminator. We denote the three discriminators as $D$ and the rest part of the proposed Expressive-VC model as generator $G$. The loss functions of $G$ and $D$ can be described as:
\vspace{-6pt}
\begin{equation}
\mathcal{L}_{G}(Y, \hat{Y})=\mathcal{L}_{adv_g}(Y, \hat{Y})+\mathcal{L}_{fm}(Y, \hat{Y})+\mathcal{L}_{stft}(Y, \hat{Y}),
\vspace{-6pt}
\end{equation}
\vspace{-6pt}
\begin{equation}
\mathcal{L}_{D}(Y, \hat{Y})=\mathcal{L}_{adv_d}(Y, \hat{Y}),
\vspace{-6pt}
\end{equation}
where $Y$ and $\hat{Y}$ are ground-truth and predicted waveform. $\mathcal{L}_{adv_g}$ and $\mathcal{L}_{adv_D}$ are the adversarial loss of the generator and discriminator respectively. Besides, the feature matching loss $\mathcal{L}_{fm}$~\cite{DBLP:conf/nips/SalimansGZCRCC16} and the multi-resolution STFT
loss $\mathcal{L}_{stft}$ ~\cite{DBLP:conf/icassp/TakakiNWY19} are also adopted. As mentioned in Section~\ref{sec:training}, except predicting waveform $\hat{Y}_{f}$ from fusion content $H_{f}$, $H_{w}$ is also used to directly reconstruct waveform $\hat{Y_{w}}$. The overall objective function is described as
\vspace{-6pt}
\begin{equation}
\mathcal{L}_{total_G}(Y, \hat{Y}_{f}, \hat{Y}_{w})=\mathcal{L}_G(Y, \hat{Y}_{f})+\mathcal{L}_G(Y, \hat{Y}_{w}),
\vspace{-6pt}
\end{equation}
\begin{equation}
\mathcal{L}_{total_D}(Y, \hat{Y}_{f}, \hat{Y}_{w})=\mathcal{L}_{a d v_d}(Y, \hat{Y}_{f})+\mathcal{L}_{adv_d}(Y, \hat{Y}_{w}).
\vspace{-6pt}
\end{equation}

\vspace{-8pt}
\subsection{Training Strategy for Forcing Feature Fusion}
\vspace{-4pt}
\label{sec:training}
% In recognition-synthesis-based approaches, models can stably reconstruct mel spectrograms with good intelligibility, showing that BNFs can well represent the semantic information in waves. Since a large amount of information is already contained in the BNFs, the model tends to use the BNFs directly, rather than the learned LFs from the content encoder, causing insufficient training of the content encoder. Therefore, the weights obtained by the fusion module will be heavily biased towards BNFs, degenerating the model to a recognition-synthesis-based approach.

% To alleviate this problem, we must force the content encoder to learn more information and balance the weights between the two features. An extra linguistic loss is added to the total training loss, which is a set of losses that are identical to the basic training loss, forcing the model to reconstruct waveform only from LFs, which in turn improves the quality of the features extracted by the content encoder. Due to the limitation of computing resources, these two sets of losses cannot be calculated at the same time. 
% Our solution is to calculate these two sets of losses, in turn, that is, select different features for the decoder for each training step, whether using the fused features or the standalone LFs.

Ideally, the fusion module should learn to obtain linguistic information from $H_{b}$ while extracting para-linguistic information from $H_{w}$ that $H_{b}$ can not well represent. However, in practice, since BNFs are more related to linguistic information than the perturbed waveform, learning linguistic information from BNF encoder is much easier than that from the Perturbed-wav encoder. Consequently, the fusion module tends to only focus on the linguistic information $H_b$ extracted from BNFs, causing the failure of the fusion module and the Perturbed-wav encoder. To encourage better feature fusion, the convergence speed and content extraction ability of the Perturbed-wav encoder need to be particularly strengthened during training. As shown in the red arrow in  Fig.~\ref{fig:model}(a), bypassing the fusion module, $H_{w}$ and $H_{p}$ are directly added and then fed into the decoder for waveform reconstruction. Through this auxiliary training, the Perturbed-wav encoder can be directly guided by waveform reconstruction and optimized faster. Finally with this training trick and the prosody encoder provided query, the fusion module can perform more reasonable fusion between $H_{b}$ and $H_{w}$.

\vspace{-15pt}
\section{Experiments}
\vspace{-8pt}
\label{sec:pagestyle}

\begin{figure*}[!htbp]
    % \label{fig:pitch}
    \centering 
  \begin{minipage}[t]{0.30\linewidth}
    \centering    
% \subfloat[content encoder]{\includegraphics[trim=20 0 20 0,scale=0.7]{IEEEtran/fig/content.pdf}}
\subfloat[Normal speech]{\includegraphics[clip,scale=0.17]{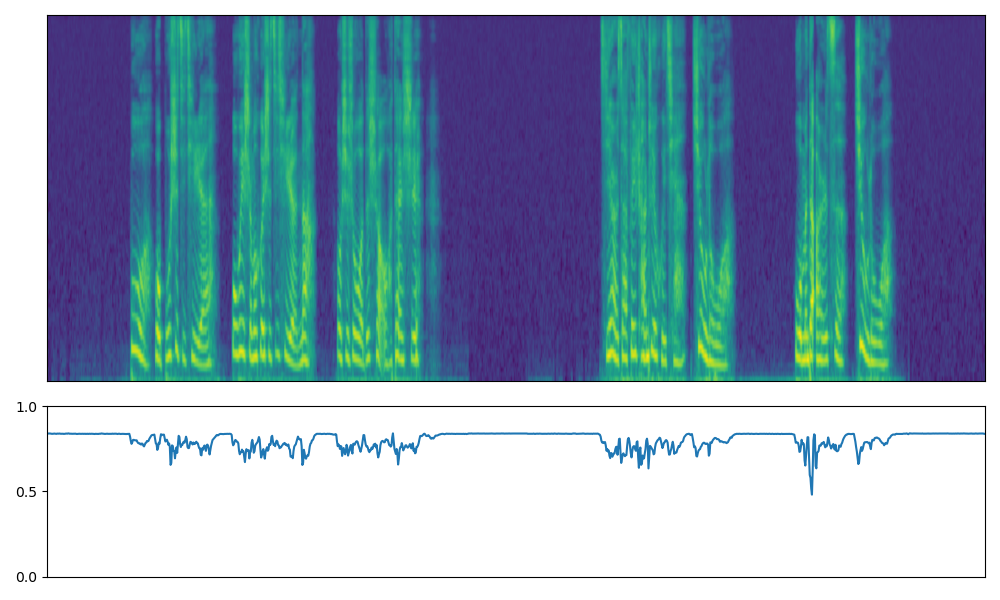}}
    % \label{fig:side:a}
  \end{minipage}%
  \begin{minipage}[t]{0.30\linewidth}
    \centering
\subfloat[Shouting]{\includegraphics[clip,scale=0.17]{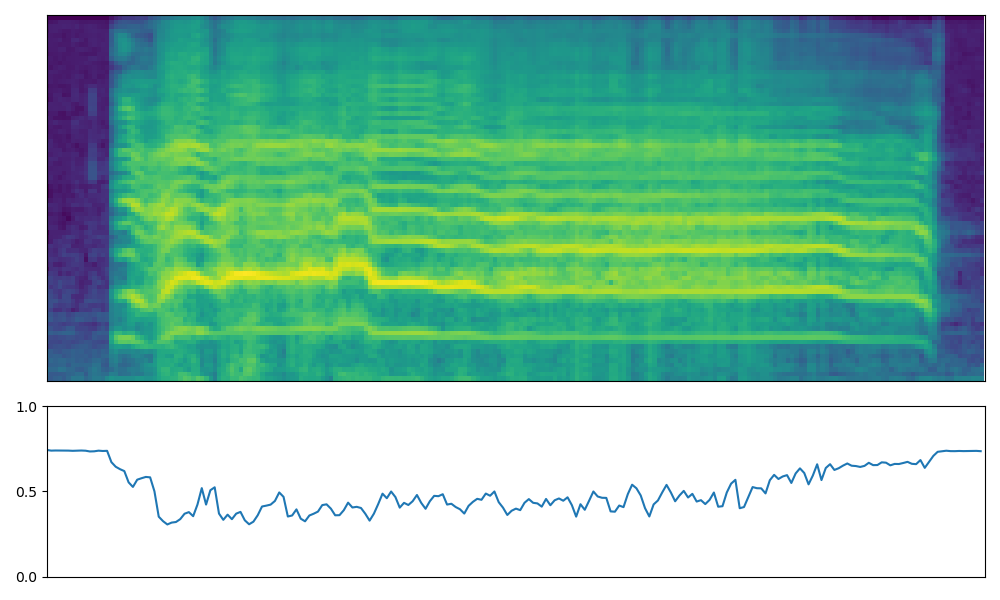}}
    % \caption{cap b}
    % \label{fig:side:b}
  \end{minipage}
    \begin{minipage}[t]{0.30\linewidth}
    \centering
\subfloat[Gasping]{\includegraphics[clip,scale=0.17]{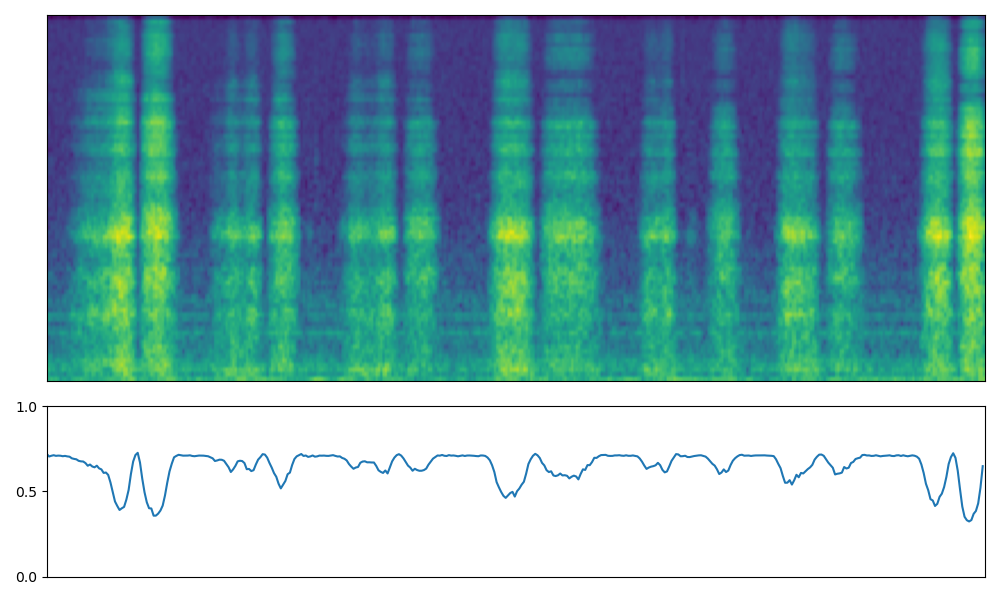}}
    % \caption{cap b}
    % \label{fig:side:b}
  \end{minipage}
  \vspace{-8pt} 
     \caption{Spectrograms (the first row) of normal speech, shouting and gasping with their corresponding BNF weight trajectory (the second row).}
    \label{fig:weightmatrix}
    \vspace{-18pt} 
\end{figure*}
% \begin{figure}[!htb]
%  \vspace{-0.2cm}
%   \begin{minipage}[t]{0.5\linewidth}
%     \centering    
% \subfloat[weight matrix]{\includegraphics[trim=15 0 15 0,scale=0.35]{figs/E01_f_3.png}}
%     % \caption{cap b}
%     \label{fig:side:b}
%   \end{minipage}
%      \caption{weight matrix of features fused by a four-head attention}\vspace{-0.3cm}
%     \label{fig:weightmatrix} 
%\end{figure}
\vspace{-0.1cm}

\subsection{Experimental setup}
\vspace{-5pt}
In the experiments, all testing VC models are trained on an internal Mandarin corpus, containing about 100K neutral utterances and 60K expressive utterances uttered by 230 speakers. One male and one female speakers are reserved as the target speakers for voice conversion tests. A set of 30 speech recordings, including typical reading (non-expressive), expressive and non-verbal clips, are used as source data, 10 utterances for each category. The selected recordings  are converted to the two target speakers using the proposed model and all the comparison models to further perform evaluations. All the speech utterances are resampled to 24 kHz. For perturbation methods, we conducted formant shifting, pitch randomization, and random frequency shaping to the waveform with the same perturbation coefficient as NANSY~\cite{DBLP:conf/nips/ChoiLKLHL21}. Besides, speed augmentation is adopted to enrich prosody diversity~\cite{DBLP:conf/icassp/ZhaoLSWKTM22}, using a random multiplier of 1.1-1.5. During training, augmented and original waveforms are fed to the VC model alternatively.
Mel spectrum, pitch, and energy are computed with 50ms frame length and 10ms hop size. The ASR system for BNF extraction is a conformer-based model trained on a Mandarin ASR corpus Wenetspeech~\cite{DBLP:conf/icassp/ZhangLGSYXXBCZW22}, implemented by WeNet toolkit~\cite{DBLP:conf/interspeech/YaoWWZYYPCXL21}. In our implementation, the BNF encoder consists of two convolution layers, each followed by layer normalization. The Perturbed-wav encoder consists of convolution layers with four strides, downsampled by a factor of 6, 5, 5, and 2, each of which is also followed by a layer normalization. The architecture and hyper-parameters of the prosody encoder, decoder, and discriminator follow the origin configuration in~\cite{DBLP:journals/corr/abs-2206-07569}.

To validate the performance of the proposed model in highly expressive voice conversion, \textbf{BNF-VC}~\cite{DBLP:conf/interspeech/WangZYLDXGCL21}, \textbf{Perturb-VC}~\cite{DBLP:journals/corr/abs-2206-07569}, and \textbf{AGAIN-VC}~\cite{DBLP:conf/icassp/ChenWWL21} are used as our comparison systems, all implemented using the same training data described above. BNF-VC, based on the BNF framework and helped with explicit prosody modeling, is a good representative of balancing intelligibility, expressiveness and speaker similarity. Perturb-VC is a newly perturbation based end-to-end model while AGAIN-VC is another popular approach that has open-source code.\footnote{https://github.com/KimythAnly/AGAIN-VC}

\vspace{-8pt}
\subsection{Subjective Evaluation}
\vspace{-6pt}
We conduct Mean Opinion Score (MOS) tests to evaluate the naturalness and speaker similarity of different models. Since this paper aims to perform highly expressive voice conversion, the naturalness metric considers the consistency between source speech and converted speech in terms of expressiveness and pronunciation. Higher naturalness MOS score means converted speech can better maintain the expressiveness of the source speech. In both MOS tests, there are 20 listeners participated. Particularly for speaker similarity test, we use target speaker's real recording as reference. We recommend the readers listen to our samples\footnote{Demo: https://nzqian.github.io/Expressive-VC.github.io/}.

% the degree of preservation of the expressiveness and style of the source speech and the number of defects in the pronunciation. Since the expressiveness is not the higher the better, but the closer it is to the source speech, the better, all the samples are of high expressiveness, and a higher MOS score proves that the model's expressiveness is better.

\noindent\textbf{Speech Naturalness Evaluation}
The results shown in Table~\ref{tab:mos} indicate that our proposed Expressive-VC can achieve the best performance in speech naturalness. Specifically, in the non-expressive scenario, BNF-VC gets a similar score to Expressive-VC, which shows that BNF can well represent content information of speech in this scenario. For expressive validation, all comparison systems show performance degradation. Expressive-VC gets a higher MOS score which shows that Expressive-VC can capture more rich content with para-linguistic information. Particularly for non-verbal cases, Expressive-VC has obvious superiority in naturalness and the MOS score remains at the same level with expressive and non-expressive cases.

% in the normal scenario, the scores of the proposed model are not significantly different compared to the BN model, due to the fact that the asr model is able to obtain robust content representation for speech in normal scenes. For emotional and stylistic and highly expressive data scenarios, the proposed model scores significantly better than the BN model. We believe that, on the one hand, the ASR training data lacks highly expressive data, and on the other hand, since the ASR system is trained on phonemes, pronunciation fragments that cannot be described by words or phonemes, such as coughing and wheezing, will lead to impairment of the content representation extracted by the ASR system itself. The proposed model, however, contains an end-to-end modeling approach that is able to directly model generic pronunciation representations. The proposed approach has significant advantages over the Disentangling-based system in all scenarios. This is due to the fact that the input of the proposed model contains the content representation extracted by the ASR system, which is more robust compared to the content representation extracted by the Disentangling-based system.

\noindent\textbf{Speaker Similarity Evaluation}
The results of MOS tests in terms of speaker similarity for different models are also shown in Table~\ref{tab:mos}, in which higher MOS means better performance. In the non-expressive scenario, BNF-VC achieves higher speaker similarity than other systems. We also notice that speaker similarity still remains at a high level for non-verbal cases for Expressive-VC. In expressive and non-verbal scenarios, compared with BNF-VC, Perturb-VC, and AGAIN-VC, the proposed method achieves better performance in speaker similarity. Considering the overall performance of speaker similarity and the superiority of the proposed method in the naturalness, Expressive-VC shows better performance in the highly expressive voice conversion.

% For speaker similarity, compared to the BN system, the proposed model has little difference in scores in normal scenes, but scores higher in highly expressive scenes. For the BN system at highly expressive scenes, the content information is impaired, resulting in some content information leaking into the timbre space. Due to the disentangling-based approach, there will be incomplete disentangling of content information and timbre information, which brings loss of timbre modeling.
\begin{table}[]
\centering
 \caption{Pearson correlation in energy and lf0.}
 \vspace{-8pt}
\setlength{\tabcolsep}{0.6mm}
 \label{tab:corrcoef}
  \resizebox{\linewidth}{!}{
\begin{tabular}{@{}c|cccc@{}}
\toprule
    & Expressive-VC & BNF-VC~\cite{DBLP:conf/interspeech/WangZYLDXGCL21} & Perturb-VC~\cite{DBLP:journals/corr/abs-2206-07569} & AGAIN-VC~\cite{DBLP:conf/icassp/ChenWWL21} \\ \midrule
LF0 $\uparrow$& \textbf{0.754}    & 0.741   & 0.625      & 0.368    \\ 
Energy $\uparrow$& \textbf{0.985}    & 0.977   & 0.982      & 0.903    \\ 
\bottomrule
\end{tabular}
}
\vspace{-20pt}
\end{table}

\noindent\textbf{Ablation Study}
To investigate the importance of our proposed methods in Expressive-VC, three ablation systems were obtained by dropping the fusion module, auxiliary training loss ($\mathcal{L}_{G_c},\mathcal{L}_{D_c}$), and speed augmentation, referred to as \textit{-Fusion Module}, \textit{-$\mathcal{L}_{G_c},\mathcal{L}_{D_c}$}, and \textit{-Speed Aug}. Note that when dropping the fusion module, the fusion of features is performed by direct concatenation. As shown in Table~\ref{tab:mos}, dropping these methods brings obvious performance degradation in terms of speech naturalness and speaker similarity. Specifically, without the fusion module, the concatenation of two features cannot be dynamically adjusted, leading to performance degradation. Besides, when $\mathcal{L}_{G_c}$ and $\mathcal{L}_{D_c}$ are discarded, the process of feature fusion fails, and the performance drops in both naturalness and speaker similarity. As can be seen, speech speed augmentation also contributes to the proposed system's performance.

% The experimental results are shown in the table, where -linguistic loss and -duration aug indicate that no linguistic loss and no data augmentation on duration are used, respectively, and -attention indicates that the proposed attention-based fusion algorithm is not used, but the combination of features is performed by direct splicing.

% As can be seen from the table, dropping any component leads to significant performance degradation in terms of naturalness and speaker similarity.
% Specifically, the greatest degradation is brought by not using linguistic loss, in which case the model tends to learn content representation from BN inputs and ignore input information from decoupled branches, degrading the model into a Recognition-synthesis-based model. For time-length data augmentation, the rhythmic diversity of the data is reduced, which in turn leads to a decrease in naturalness. And if the features are fused in a direct splicing way without using attention, the proportion of two features in the fused features cannot be dynamically adjusted according to the characteristics of each moment in the audio.

\vspace{-14pt}
\subsection{Objective Analysis}
\vspace{-6pt}
\textbf{Character Error Rate} We use the same pre-trained ASR model for BNF extraction to recognize the source speech, converted non-expressive and expressive speech clips. The character error rate is also reported in Table~\ref{tab:mos}. We can see that AGAIN-VC obtains the highest CER, indicating bad intelligibility. By contrast, our Expressive-VC has similar CER with BNF-VC, while both systems induce small CER increase as compared with the source speech. We also notice that the CER is much higher for expressive clips as compared with non-expressive counterparts. We believe expressive speech is more difficult to recognize by a regular ASR system. In summary, the proposed system still can maintain reasonable intelligibility. 

\noindent\textbf{Visualization on Fusion Process} To further study the process of feature fusion, as shown in Fig.~\ref{fig:weightmatrix}, three speech recordings containing normal speech, shouting, and gasping are used as source speech to obtain the mel spectrums and attention weights. Note that the weight curves vary from 0 to 1, indicating the proportion of $H_b$ in the fusion feature.
As can be seen, compared with the weight curve of normal speech in Fig.~\ref{fig:weightmatrix}(a), the weight curves of Fig.~\ref{fig:weightmatrix}(b) and Fig.~\ref{fig:weightmatrix}(c) are much lower when shouting and gasping happen, indicating that the $H_w$ extracted from the perturbed waveform is more involved in the fusion process. Moreover, the shouting in Fig.~\ref{fig:weightmatrix}(b) gradually weakens over time, corresponding to a gradual increase in the weight of $H_b$, demonstrating that the model can flexibly adjust the proportion of $H_b$ and $H_w$ in the fusion feature according to the expression and pronunciation in different times. These suggest that the two features are fused as we expected: $H_b$ from BNFs mainly contains linguistic formation, and the $H_w$ from perturbed waveform  provides additional para-linguistic information.

%  We can intuitively see the change in the feature weights of BNFs and LFs in the fused features over time, by using the modified version of the attention mechanism. Fig.~\ref{fig:weightmatrix} shows the spectrogram of three converted speech and their corresponding attention weight, which are speech in normal tones, shouting, and gasping. The weight curves in the figures vary between 0-1, indicating the proportion of BNFs in the fused feature. In fig.~\ref{fig:weightmatrix}(b) and fig.~\ref{fig:weightmatrix}(c), the weight is significantly lower than the normal tones when shouting and gasping, suggesting that more LFs are used when the model is faced with highly expressive scenes. The shout in fig.~\ref{fig:weightmatrix}(b) gradually weakens over time, corresponding to a gradual decrease in the proportion of LFs, demonstrating that the model can flexibly adjust the proportion of the two features in the fused features according to the strength of the sentence expression. This proves that the two features are fused in the same way as we expected: BNFs contain mainly semantic information, and the LFs contain additional expressive information. 

\noindent\textbf{Pitch Correlation} 
%To further verify the expressiveness of each system, we extracted energy and pitch from 120 samples to calculate the Pearson correlation coefficients of all systems. 
To further verify the expressiveness of each system, we calculate the Pearson correlation coefficients of energy and pitch between source and converted speech of all systems. 
The higher the Pearson correlation coefficient of the model, the higher the accuracy of the predicted prosodic attributes. 
%As shown in Table~\ref{tab:corrcoef}, 
Table~\ref{tab:corrcoef} shows that1
the proposed system has the highest lf0 and energy correlations. It illustrates that the proposed system can better maintain the expressive aspects of the source speech compared to the comparison systems.

\vspace{-14pt}
\section{Conclusions}
\vspace{-10pt}
\label{sec:typestyle}

In this paper, we propose Expressive-VC for highly expressive voice conversion. This task is challenging due to the difficulty of maintaining both the linguistic and para-linguistic information in the source speech while achieving high-quality voice conversion with target speaker's timbre. To this end, multiple feature fusion is proposed in a specifically designed network structure, leveraging the advances from both bottleneck feature approach and the signal perturbation approach. Extensive experiments show that Expressive-VC achieves superior performance in highly expressive voice conversion tasks, including non-verbal sound conversion.
% Specifically, the fusion content generated by the proposed fusion process mainly obtains linguistic information from BNF and fetches the para-linguistic and non-verbal information from perturbed waveform, which BNF can not represent correctly.
% an attention-based 
% content representations extracted from BNF and perturbed waveform. In feature fusion process, the fusion content Using the proposed fusion module based on scaled dot-product attention, Expressive-VC combines the stable linguistic information in BNF and the para-linguistic information extracted from perturbed source waveform into a fused content representation. Speaker-independent information in the source speech can be effectively preserved in this way, thus ables the model to perform expressive voice conversion. 

\vfill
\pagebreak

% \section{REFERENCES}
% \label{sec:refs}

\bibliographystyle{IEEEbib}
\bibliography{strings,refs}

\end{document}